\documentstyle[twocolumn,aps,prb,epsf]{revtex}
\begin{document}
\newcommand{\beq}{\begin{equation}}
\newcommand{\eeq}{\end{equation}}

\title{Microwave absorption in $s$- and $d$-wave disordered superconductors}

\author{Mai Suan Li}

\address{Institute of Physics, Polish Academy of Sciences,
Al. Lotnikow 32/46, 02-668 Warsaw, Poland}

\address{
\centering{
\medskip\em
{}~\\
\begin{minipage}{14cm}
We model $s$- and $d$-wave ceramic superconductors 
with a three-dimensional lattice of randomly distributed
0 and $\pi$ Josephson junctions with finite self-inductance. 
The field and temperature dependences of the microwave absoption  
are obtained 
by solving the corresponding Langevin dynamical equations. 
We find that  at magnetic field $H=0$ the microwave absoption of the $s$-wave samples,
when plotted against the field, has a minimum at any temperature.  
In the case of $d$-wave superconductors one has
a peak at $H=0$ in the temperature region where the paramagnetic Meissner effect is
observable. These results
agree with experiments.
The dependence of the microwave absorption on the screening strength was found to be
nontrivial due to the crossover from the weak to the strong screening regime.
{}~\\
{}~
\end{minipage}
}}

\maketitle



\section{Introduction}

A fundamental property of superconductivity is the Meissner effect, i.e.
the occurence of flux expulsion below the superconducting transition temperature
and the resulting diamagnetic response to the external magnetic field. Contrary
to this behavior a paramagnetic signal was observed in certain ceramic superconductors
upon cooling in low enough fields (smaller than 0.1 mT) \cite{Svelindh,Braunish}.
This effect is now referred to as the paramagnetic Meissner effect (PME) or
the  Wohlleben effect. The nature of the unusual paramagnetic behavior may be related to
the appearance of the spontaneous supercurrents (or of orbital moments)
\cite{Kusmartsev}. The latter occur due to the existence
of $\pi$-junctions characterized by the negative Josephson couplings
\cite{Kusmartsev,Bulaevskii}.
Furthermore, Sigrist and Rice \cite{Sirgist,Sirgist1} argued that the PME in the high-$T_c$
superconductors is a consequence of the
intrinsic unconventional pairing symmetry
of the HTCS of $d_{x^2-y^2}$ type.\cite{wollman}
In fact, the PME is succesfully reproduced in a single
loop model\cite{Sirgist} as well as in a model of interacting junction-loops
\cite{Dominguez,KawLi}. The latter model incorporates  
a network of Josephson junctions with a random concentration of $\pi$-junctions.
The magnetic screening is taken into account in
both of the single and multi $\pi$-junction systems.

The mechanism of the PME based on the $d$-wave symmetry of the order parameter
remains ambiguous because it is not clear why this effect could not be
observed in many ceramic materials. Furthermore, the paramagnetic response
has been seen even in the conventional Nb  \cite{Thompson,Kostic,Pust},
 Al \cite{Geim} superconductors and the Nb-AlO$_x$-Nb tunnel junctions
\cite{Barbara}.
In order to explain the PME in terms of conventional superconductivity
one can employ the idea of the flux compression inside of a sample. Such
phenomenon becomes possible in the presence of the inhomogeneities\cite{Larkin}
or of the sample boundary\cite{Moshchalkov}.  Auletta {\em et al} \cite{Auletta}
have also observed the PME in the model of special geometry involving only 0-junctions. 
In our opinion, the PME in this model is of the dynamical nature but not the
equillibrium effect as in the $d$-wave model \cite{KawLi}.
Thus the intrinsic mechanism leading to the PME is still under
debate\cite{Geim,Sigrist2}.

One of the most valuable tools to distinguish between the $s-$ and $d$-wave
symmetry is the study of the microwave absorption (MWA) \cite{Sigrist2}.
In fact, Braunish {\em et al}  \cite{Braunish} have found a nontrivial field dependence
of the MWA in samples which display the PME.
The MWA has a peak at $H=0$, when plotted against $H$,
whereas for $s$-wave superconductors it has a conventional minimum. Based on a hysteresis in
the $M-H$ space ($M$ is a magnetization) Sigrist and Rice have shown that
for the one-loop model the peak at $H=0$ would be observed if the 
dimensionless self-inductance,
$\tilde{L}$, exceeds some borderline value $\tilde{L}^*=1$.
The relation between $\tilde{L}$ and
the inductance, $L$ is as follows
\begin{equation}
 \tilde{L} \; \; = \; \; \frac{2\pi I_c}{c\Phi_o} L \; ,
\end{equation}
where $\Phi_0$ and $I_c$ are 
the flux quantum and the critical current, respectively.

It should be noted that Dominguez {\em et al} \cite{Dominguez1} have 
qualitatively reproduced the experimental
findings for the MWA using the multi-loop model. Their results are, however, 
restricted to the two-dimensional system. More important, the
question about the borderline value $L^*$
above which the nontrivial field dependence of
the MWA may occur in the $d$-wave interacting loops model was not
studied. Also the role of temperature and
of the screening have not been explored 
(the screening, for example, plays a key role
in explaing experiments on the aging effect in ceramic 
superconductors \cite{Li1}).

Since the underlying mechanism for the PME is still lacking, a
careful study of MWA may shed  some light on this problem.
In this paper we study in detail the MWA in the three-dimensional 
system which is
more relevant to experimental situations than the two-dimensional one.
Integrating the corresponding Langevin equations we have made three new
observations. First, 
contrary to the one-loop model the MWA in the system with randomly
distributed $\pi$-junctions has a non-trivial field dependence for any value of
$\tilde{L}$. In other words, $\tilde{L}^*=0$ for the multi-loop model. 
Second, the peak 
at $H=0$ is
found to disappear for $T > T^*$, where $T^*$ is a boderline temperature below
which the PME is observable.  $T^*$ was found to grow as the screening is
lowered.
The third observation is  
that  for both $s$- and $d$-wave ceramics the MWA decreases with screening not
monotonically but it has a minimum
at $\tilde{L }\approx 1$. Such a behavior is related to a change in time and length scales 
when one goes from the weak to the strong screening limit.

\section{Model}

We neglect the charging effects of the grains and
consider the following Hamiltonian\cite{Dominguez,KawLi}
\begin{eqnarray}
{\cal H} = - \sum _{<ij>} J_{ij}\cos (\theta _i-\theta _j-A_{ij})+ \nonumber\\
\frac {1}{2{ L}} \sum _p (\Phi_p - \Phi_p^{ext})^2, \nonumber\\
\Phi_p \; \; = \; \; \frac{\phi_0}{2\pi} \sum_{<ij>}^{p} A_{ij} \; , \;
A_{ij} \; = \; \frac{2\pi}{\phi_0} \int_{i}^{j} \, \vec{A}(\vec{r})
d\vec{r} \; \; ,
\end{eqnarray}
where $\theta _i$ is the phase of the condensate of the grain
at the $i$-th site of a simple cubic lattice,
$\vec A$ is the fluctuating gauge potential at each link
of the lattice,
$\phi _0$ denotes the flux quantum,
$J_{ij}$ denotes the Josephson coupling
between the $i$-th and $j$-th grains,
$ L$ is the self-inductance of a loop (an elementary plaquette),
while the mutual inductance between different loops
is neglected.
The first sum is taken over all nearest-neighbor pairs and the
second sum is taken over all elementary plaquettes on the lattice.
Fluctuating  variables to be summed over are the phase variables,
$\theta _i$, at each site and the gauge variables, $A_{ij}$, at each
link. $\Phi_p$ is the total magnetic flux threading through the
$p$-th plaquette, whereas $\Phi_p^{ext}$ is the flux due to an
external magnetic field applied along the $z$-direction,
\begin{equation}
\Phi_p^{ext} = \left\{ \begin{array}{ll}
                   HS \; \;  & \mbox{if $p$ is on the $<xy>$ plane}\\
                   0  & \mbox{otherwise} \; \; ,
                        \end{array}
                  \right.
\end{equation}
where $S$ denotes the area of an elementary plaquette.
For the $d$-wave superconductors
we assume $J_{ij}$ to be an independent random variable
taking the values $J$ or $-J$ with equal probability ($\pm J$ or bimodal
distribution), each representing 0 and $\pi$ junctions.
In the case of $s$-wave ceramics $J_{ij}$ is always positive but distributed
uniformly between 0 and 2$J$.

It should be noted that model (2) is adequate to describe many
dynamical phenomena related to the PME such as 
the compesation effect \cite{Li}, the aging phenomenon \cite{Li1}, 
the effect
of applied electric fields in the apparent critical current \cite{DWJ}
and the ac resistivity \cite{Li2}.

In order to study the MWA we have to calculate the linear response to an
external electromagnetic field. Using the relation between the
MWA and the conductivity and the Kubo formula \cite{Kubo} one can show that this
response is proportional to a voltage -- voltage correlation function.
Integrating over all of frequencies of the electromagnetic field we obtain the following
expression for the MWA
\begin{equation}
\Omega \; \; = \; \; \frac{4\pi}{cnRT} \sum_i \; < V_i^2 > \; \; ,
\end{equation}
where  $ < V_i^2 >$ is a mean value of the square of the voltage induced by the
thermal noise on each junction, $n$ is a light refraction coefficient
and $R$ is the normal resistance of the links.

To calculate $V_i$ we
model the current flowing between two grains with the resistively shunted
junction (RSJ) model,\cite{Dominguez,Joseph2} which gives
the following  dynamical equations:
\begin{eqnarray}
\frac{\hbar}{2eR}\frac{d\theta_\mu({\bf n})}{dt} \; = \;
-\frac{2e}{\hbar}J_\mu({\bf n})\sin\theta_\mu({\bf n})
\nonumber\\
-\frac{\hbar}{2e{\cal L}}\Delta_\nu^{-}\left[
\Delta_\nu^{+}\theta_{\mu}({\bf n})-\Delta_\mu^{+}\theta_\nu({\bf n})\right]
 -\eta_\mu({\bf n},t) \, \, .
\end{eqnarray}
Here we have redefined notation:
the site of each grain is at position ${\bf n}=(n_x,n_y,n_z)$
(i.e. $i\equiv{\bf n}$); the lattice directions are
$\mu={\hat{\bf x}}, {\hat{\bf y}}, {\hat{\bf z}}$;
the link variables are between sites ${\bf n}$ and ${\bf n}+\mu$
(i.e. link $ij$  $\equiv$ link ${\bf n},\mu$);
and the plaquettes $p$ are defined by the site ${\bf n}$ and
the normal direction $\mu$ (i.e plaquette $p$ $\equiv$ plaquette
${\bf n},\mu$, for example the plaquette ${\bf n}, {\hat{\bf z}}$ is
centered at position ${\bf n}+({\hat{\bf x}}+{\hat{\bf y}})/2$).
The forward difference operator $\Delta_{\mu}^{+}\theta_\nu({\bf n})=
\theta_\nu({\bf n}+\mu)-\theta_\nu({\bf n})$
and the backward operator  $\Delta_{\mu}^{-}\theta_\nu({\bf n})=\theta_\nu({\bf n})-
\theta_\nu({\bf n}-\mu)$.
The Langevin noise
current $\eta_{\mu}({\bf n},t)$ has Gaussian correlations
\begin{equation}
\langle\eta_{\mu}({\bf n},t)\eta_{\mu '}({\bf n}',t') \; = \; 
\frac{2k_BT}{R}\delta_{\mu ,\mu'}
\delta_{{\bf n} , {\bf n'}}\delta(t-t') \; .
\end{equation}
The local voltage $V_i$ is then given by
\begin{equation}
V_i \; \; = \; \; \frac{d\theta_i}{dt} \; \; .
\end{equation}

Eq. (5) describes the overdamped dynamics. We have tried to include the inertia
(capacitive) terms but the results do not change  substantially  and they are neglected.

In what follows we will consider currents normalized by $I_J=2eJ/\hbar$,
time by $\tau=\phi_0/2\pi I_JR$, voltages by $RI_J$, inductance by
$\phi_0/2\pi I_J$ and temperature by $J/k_B$.	     
Then the dimensionless
MWA, $\tilde{\Omega}$ is defined as follows
\begin{equation}
\tilde{\Omega} \; \; = \; \; \frac{cnR}{4\pi} \Omega \; .
\end{equation}

\section{Results}

The system of differential equations (5) is integrated numerically by a
second order Runge-Kutta-Helfand-Greenside algorithm for stochastic
differential equations \cite{RKutta}. The time step depends on $\tilde{L}$ and
is equal to $\Delta t = 0.1\tau _J$ and 
$\Delta t = 0.1\tau _J \times \tilde{L}$ for  $\tilde{L} > 1$ and $\tilde{L} < 1$,
respectively.
We conider the system size $l=8$ (we have made
some test runs for $l=12$ and found that the finite size effects are not substantial).
The temporal averages are taken over a time of  $10^5 \tau _J$ after a transient time
of $25000 \tau _J$. 
The free boundary conditions are implemented
because the magnetization always vanishes for the periodic boundary conditions
\cite{Dominguez,KawLi}

\begin{figure}
\epsfxsize=3.2in
\centerline{\epsffile{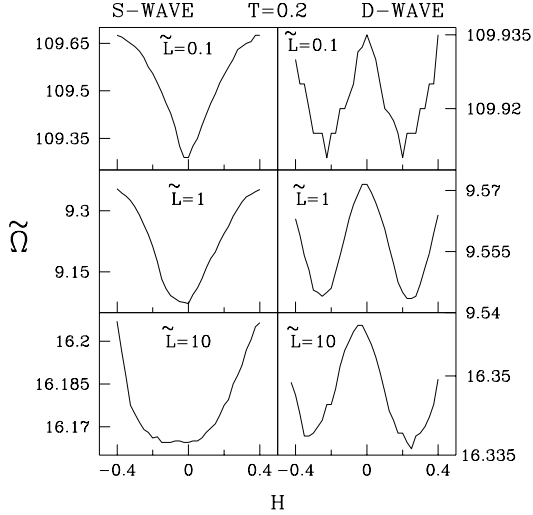}}
\vspace{0.03in}
\caption{The field dependence of $\Omega$ for $s$- (left panel) and $d$-wave
(right panel) ceramic superconductors. We choose $T=0.2$ and
$\tilde{L}=0.1, 1$ and 10.
The results are averaged over
20  samples.}
\end{figure}

\begin{figure}
\epsfxsize=3.2in
\centerline{\epsffile{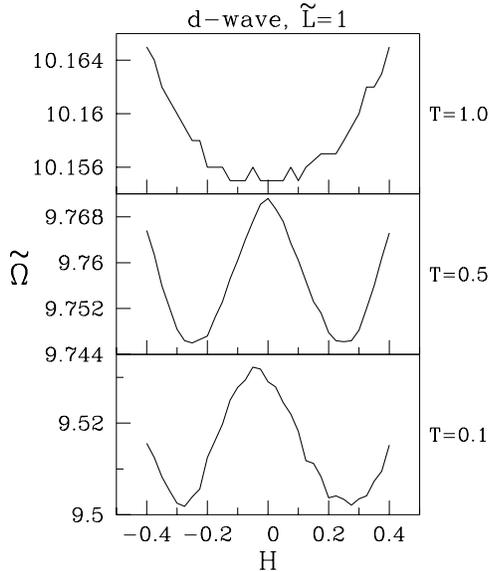}}
\vspace{0.07in}
\caption{The field dependence of the MWA for the $d$-wave samples.
We took $l=8$, $\tilde{L}=1$ and $T=0.1, 0.5$ and 1. The results are
averaged over 10 -20 samples.}
\end{figure}

Fig. 1 shows the field dependence of the MWA for $T$=0.2 and for various values
of $\tilde{L}$. In the case of $s$-wave superconductors we have the 
standard minimum
at $H=0$ for any value of inductance. It is also true for any $T$. As expected, 
$\tilde{\Omega} \sim H^2$ at weak fields. For the $d$-wave samples $\Omega$ has
the unconventional peak at $H=0$. Contrary to the one-loop model
\cite{Sirgist} such peak is seen not only for $\tilde{L} > 1$
but also for $\tilde{L} \le 1$. 
It should be noted that the height of the peak is very small
(($\tilde{\Omega } (H=0) - \tilde{\Omega }_{min})/\tilde{\Omega }_{min}$ 
is of order of $10^{-3}$).
This is in qualitative agreement with experimental findings \cite{Braunish}
that the peak should be low.

In our model (2) the temperature dependence of the critical current is neglected.
However, one can show that the dimensionless temperature $T$ chosen in Fig. 1 and
in all of the figures presented below corresponds to the relevant to experiments real 
temperature, $T_R$. In fact, the critical current depends not only on temperature but 
also on 
conditions under which samples were prepared.
The typical value of the cirtical current density for 
ceramic superconductors
is $ \sim 10^6 A/m^2$ ( see, for example, Ref. \onlinecite{Current}). 
Since the typical size of grains is about $1\mu m$
we have the critical current $I_c \sim 10^{-6}A$.
Using $T_R=JT/k_B=\hbar I_cT/2ek_B$ one obtains $T_R/T \sim 100K$.
Clearly, our dimensionless $T$ correctly desribes the experimental values of temperature
\cite{Braunish}.

\begin{figure}
\epsfxsize=3.2in
\centerline{\epsffile{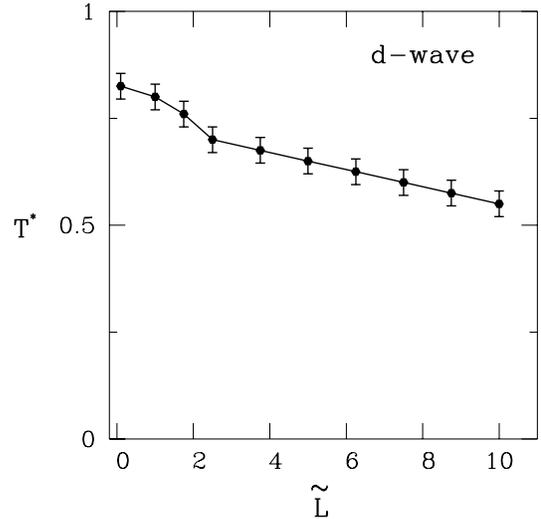}}
\caption{The inductance dependence of $T^*$ for $d$-wave superconductors.}
\end{figure}

It is known that the random $\pi$-junction model (2) 
displays a phase transition to a
so called chiral glass.\cite{KawLi1}
The frustration effect due to the random distribution of $\pi$ junctions
leads to a glass state of quenched-in ``chiralities'', which are
local loop supercurrents circulating over grains and
carrying a half-quantum of flux.
Evidence of this transition has been related to measurements
of the nonlinear ac magnetic susceptibility.\cite{Matsuura1}
The question we ask is if there is any correlation between the existence of the
chiral glass phase and the anomalous behavior of $\Omega$. As shown in Ref. 
\onlinecite{KawLi1}, the chiral glass disappears for $\tilde{L} > \tilde{L}_c$,
where $\tilde{L}_c = 5 - 7$. On the other hand, the peak of $\Omega$ is
obserable for any value of $\tilde{L}$. Therefore, there is no one-to-one
correlation between the chiral glass and the nontrivial field dependence
of the MWA of the ceramic superconductors.

The field dependence of the MWA in the $d$-wave superconductors for
$\tilde{L}=1$ and various values of $T$ is shown in Fig. 2.
At low $T$'s the peak at $H=0$ shows up but {\em it disappears at high temperatures}.
This is our main result. Such a observation was not reported in
Ref. \onlinecite{Dominguez1}.  Qualitatively, 
above some borderline temperature, $T^*$ the frustration effect
becomes less important and the $d$-wave system should behave like 
the $s$-wave one. 

Fig. 3 shows the dependence
of $T^*$ on $\tilde{L}$. The question we ask now is why $T^*$ decreases
with $\tilde{L}$.  To answer this question we study the dependence of
the roughness of the energy landscape on the screening. Since the number of 
energy local  minima should grow with the number of grains (or of spins) exponentially
\cite{Binder} we restrict our calculations to small system sizes. We took $l=6$ and
search for the local minima by the annealing procedure at $T>0$ and then by the
quenching at $T=0$.
The histogram, $P(E)$, collected from local minima which are reached from
10000 starting configuations is shown in Fig. 4. Obviously, the
energy landscape for $\tilde{L}$=1 is more rugged compared to the
$\tilde{L}$=10 case.

\begin{figure}
\epsfxsize=3.2in
\centerline{\epsffile{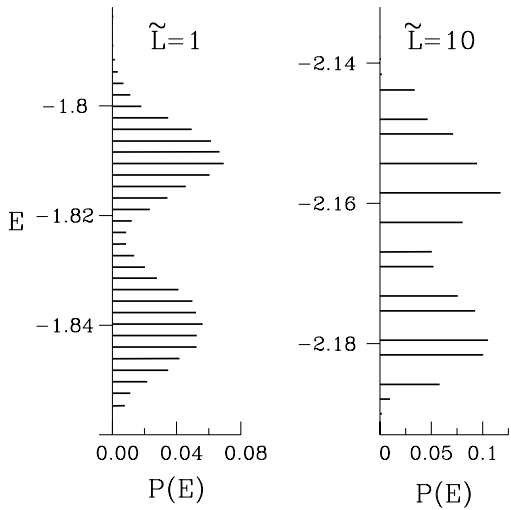}}
\caption{The energy local minima histogram, $P(E)$, for $\tilde{L}=1$ and 10. We
choose the system size $l=6$. The energy bin used for
collecting the histogram is equal to 0.002.
The local minima are obtained from 10000 starting configurations
by annealing at $T>0$ and quenching at $T=0$.}
\end{figure}

In order to characterize the roughness of the energy landscape we introduce
the parameter $\delta$,
\begin{equation}
\delta \; \; = \; \; \frac{\sqrt{<E_{lm}^2> - <E_{lm}>^2}}{<E_{lm}>} \; ,
\end{equation}
where $E_{lm}$ denotes the energy at local minima, $<...>$ means
averaging over all minima studied.
For the results presented in Fig. 4 we have $\delta=0.007$ and 0.012 for
$\tilde{L}=10$ and 1, respectively. So the larger is screening the smaller
roughness of the energy landscape. The difference between $s$- and $d$-wave
ceramics becomes, therefore, less and less pronounced as the screening
increased and $T^*$ should go down with $\tilde{L}$.

In order to understand the nature of $T^*$
we calculate the field cooled (FC) and zero field cooled (ZFC)
magnetization. In our model the magnetization is defined as follows
\begin{equation}
M \; \; = \; \; < \; \frac{1}{l(l-1)^2} \, \sum_p \frac{ \Phi_p^z}{\Phi_0} \; >
- \frac{HS}{\Phi_0} \; ,
\end{equation}
where $\Phi_p^z$ is the flux in $xy$-plane, $< ...>$ denotes the thermal and disorder average.
In the FC runs, the temperature is lowered stepwise under a constant field. At each
temperature, typically $10^5$ time steps are used for thermalization and 
$4\times 10^5$ steps for averaging.
In the ZFC runs, the system is first quenched to a low temperature ($T$=0.05)
in zero field and is thermalized during $4\times 10^5$ steps. Then a static field is 
switched on and the temperature is increased stepwise under the same condition
as in the FC regime.

\begin{figure}
\epsfxsize=3.2in
\vspace{0.07in}
\centerline{\epsffile{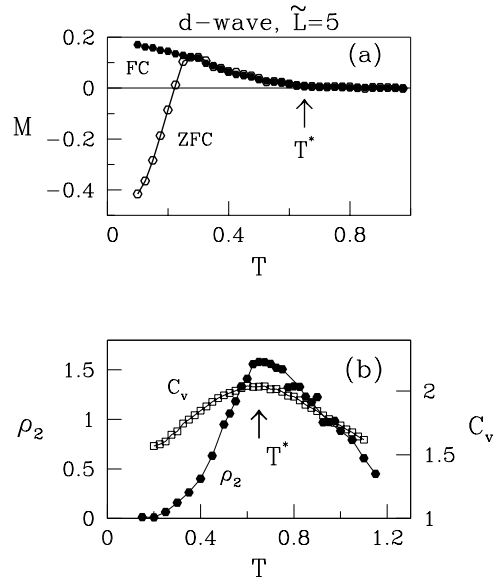}}
\vspace{0.07in}
\caption{(a)The temperature dependence of the FC and ZFC magnetization
for the $d$-wave superconductors. $\tilde{L}=5$.
(b) The same as in (a) but for $C_v$ (right-hand scale) and $\rho_2$ (left-hand scale).
The latter was computed for the frequency of the ac electric field
$\omega=0.001$ and its magnitude $I_0=0.1$ (see Ref. 23 for details).
The results are averaged over 20 - 40 samples.}
\end{figure}

Fig.5a shows the FC and ZFC magnetization for $\tilde{L}=5$
at a finite magnetic field $f=HS/\phi_0=0.1$.
We can see that $T^*$ is the temperature below which one has an
onset of positive magnetization, i.e. the paramagnetic  Meissner effect
starts to be observed.
The irreversibility point occurs at temperatures lower
than $T^*$,
and its position is dependent on the heating or cooling rate.
We identify
$T^*$  to correspond to the onset of chiral short-range order where
pair-loops begin to appear locally.
The main conclusion here is that the anomalous field dependence of the MWA is
strongly correlated with the occurence of the PME.
On the other hand, as was shown in Ref. \onlinecite{KawLi}, the PME
may appear for any value of screening, we conclude that for the multi-loop
model the borderline
value $\tilde{L}^*$ for the anomalous behavior of MWA is equal to
0.

Fig. 5b shows the temperature dependence of the specific heat, $C_v$, and
the nonlinear ac resistivity $\rho _2$ for $\tilde{L}=5$. 
$C_v$ is defined as porportional to
the energy fluctuations, $C_v=\langle(\delta E)^2\rangle/T^2$. The
definition of $\rho _2$ is given in Ref. \onlinecite{Li2}.
Clearly, $T^*$ coincides with the peak of $C_v$ and $\rho _2$.

\begin{figure}
\epsfxsize=3.2in
\centerline{\epsffile{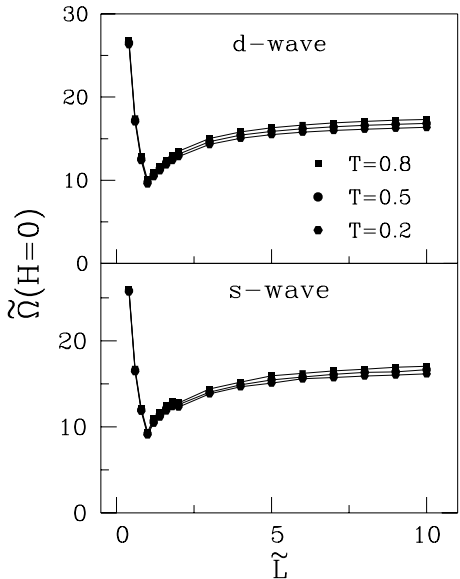}}
\vspace{0.07in}
\caption{The inductance dependence of the MWA at $H=0$ for $T=0.2$
(squares), 0.5 (hexagons) and 0.8 (circles). The system size $l=8$.
The results are averaged over 10 samples.}
\end{figure}

We now study the depedence of $\tilde{\Omega}$ on the screening. Our results are
shown in Fig. 6 for $H=0$ but the qualitative behavior is also valid 
for $H \ne 0$.
There is no appreciable difference between $s$- and $d$-wave cases. 
For a fixed value of $\tilde{L}$,  $\tilde{\Omega}(H=0)$
depends on T very weakly. The dependence on screening is more pronounced.
From  naive arguments the MWA should decreases with $\tilde{L}$  because
the screening would prevent the absorption in the bulk. Fig. 6 shows,
however, that $\tilde{\Omega}$ has a minimum at $\tilde{L}=1$. 
The anomalous dependence of MWA on  $\tilde{L}$ may be understood in the following way.
The static and dynamic properties of
the Josephson arrays are shown \cite{Rebbi,Joseph2}
to  change qualitatively if the inductance varies
from $\tilde{L}<1$ to $\tilde{L}>1$.  The attractive vortex-vortex interaction
in the weak screening regime becomes repulsive in the opposite limit. The qualitative
change in the dynamic response is related to the change of
length and time scales. Since the magnetic
screening length goes as $\lambda \sim \tilde{L}^{-1/2}$ (see Ref. \onlinecite{Joseph2}),
for $\tilde{L}<1$ $\lambda$ is larger than the grain size 
(lattice spacing in the cubic network)   
while for $\tilde{L}>1$ $\lambda$ becomes smaller than the grain size. For $\tilde{L}<1$
the relaxation time for the field is smaller than the relaxation time for the phases
whereas the opposite happens for $\tilde{L}>1$. 
The decrease of the phase relaxation time
compared to the field one  should, therefore, increase the MWA for  $\tilde{L}>1$.

\section{Conclusion}

In conclusion, experimental results of Braunish {\em et al.} \cite{Braunish}
for the MWA
can be reproduced by the XY-like model for the $d$-wave superconductor.
Although the peak of $\tilde{\Omega}$ is found to be small its
study is useful for elucidating the symmetry of the superconducting
order parameter. Within the multi-loop model
the anomalous behavior should be observable for any value of inductance
if $T < T^*$. At high temperatures there is no qualitative difference between
the $s$- and $d$-wave systems. The dependence of the MWA on the
screening strength is found to be not monotonic due to the crossover from
the weak to the strong screening regime. It would be very 
interesting to verify this
prediction experimentally. 

\acknowledgments

We thank D. Dominguez, M. Godlewski,
H. Kawamura, P. Nordblad and A. Wittlin for useful discussions.
Financial support from the Polish agency KBN
(Grant number 2P03B-146-18) is acknowledged.

\par

\end{document}